\documentclass[conference,a4paper]{IEEEtran}
\IEEEoverridecommandlockouts
\usepackage[utf8]{inputenc} 
\usepackage[T1]{fontenc}
\usepackage{url}
\usepackage{ifthen}
\usepackage{cite}
\usepackage[cmex10]{amsmath}

\usepackage{graphicx}
\usepackage{amsfonts}
\usepackage{amssymb}
\usepackage{mathrsfs}
\usepackage{enumitem}
\usepackage{subfigure}
\usepackage{mdwmath}
\usepackage{mdwtab}
\usepackage{dsfont}
\usepackage[ruled,vlined,linesnumbered]{algorithm2e} 
\newtheorem{remark}{Remark}
\newtheorem{theorem}{Theorem}
\newtheorem{lemma}{Lemma}
\newtheorem{prop}{Proposition}
\newtheorem{defi}{Definition}

\newtheorem{assump}{Assumption}

\newenvironment{iarray}{\begin{IEEEeqnarray}{rCl}}{\end{IEEEeqnarray}\ignorespacesafterend}

\def\BibTeX{{\rm B\kern-.05em{\sc i\kern-.025em b}\kern-.08em
    T\kern-.1667em\lower.7ex\hbox{E}\kern-.125emX}}
    
\begin{document}

\title{%Index Policy for Status Update Scheduling with Stochastic Packet Arrivals
Can Decentralized Status Update Achieve Universally Near-Optimal Age-of-Information in Wireless Multiaccess Channels?
}

\author{\IEEEauthorblockN{Zhiyuan Jiang$^1$, Bhaskar Krishnamachari$^2$, Sheng Zhou$^1$,  Zhisheng Niu$^1$, \IEEEmembership{Fellow,~IEEE}}
    \IEEEauthorblockA{$^1$\{zhiyuan, sheng.zhou, niuzhs\}@tsinghua.edu.cn, Tsinghua University, Beijing, China\\
        $^2$ bkrishna@usc.edu, University of Southern California, Los Angeles, USA}}

    \maketitle

\begin{abstract}
In an Internet-of-Things system where status data are collected from sensors and actuators for time-critical applications, the freshness of data is vital and can be quantified by the recently proposed age-of-information (AoI) metric. In this paper, we first consider a general scenario where multiple terminals share a common channel to transmit or receive randomly generated status packets. The optimal scheduling problem to minimize AoI is formulated as a restless multi-armed bandit problem. To solve the problem efficiently, we derive the Whittle's index in closed-form and establish the indexability thereof. Compared with existing work, we extend the index policy for AoI optimization to incorporate stochastic packet arrivals and optimal packet management (buffering the latest packet). Inspired by the index policy which has near-optimal performance but is centralized by nature, a decentralized status update scheme, i.e., the index-prioritized random access policy (IPRA), is further proposed, achieving universally near-optimal AoI performance and outperforming state-of-the-arts in the literature.
\end{abstract}


\section{Introduction}
The freshness, or timeliness, of information maintained at interested nodes is critical in many real-world applications, such as a central controller which requires time-sensitive status parameters from sensors, a cellular base station which utilizes channel state information for efficient transmissions in fast-fading scenarios, and distributed actuators where each actuator requires independent and timely inputs. With the emergence of Internet-of-Things (IoT), such information freshness is becoming increasingly important; a reasonable quantification thereof is the recently proposed age-of-information (AoI) \cite{kaul12,huang15,costa16,najm17,sun17}. Simply put, it denotes the difference between the current time and the status (maintained currently) generation time, i.e., 
\begin{equation}
    h(t) \triangleq t - \mu(t),
\end{equation} 
where $h(t)$ denotes the AoI at time $t$ and $\mu(t)$ denotes the generation time of the status maintained at destination at time $t$. In scenarios where a more up-to-date status renders an old status useless, i.e., the system status evolution is Markovian such that the current status is irrelevant with any old status given a newer status, AoI provides a compressive, and still mathematical tractable, characterization of the information freshness. 

\begin{figure}[!t]
\centering
\includegraphics[width=0.46\textwidth]{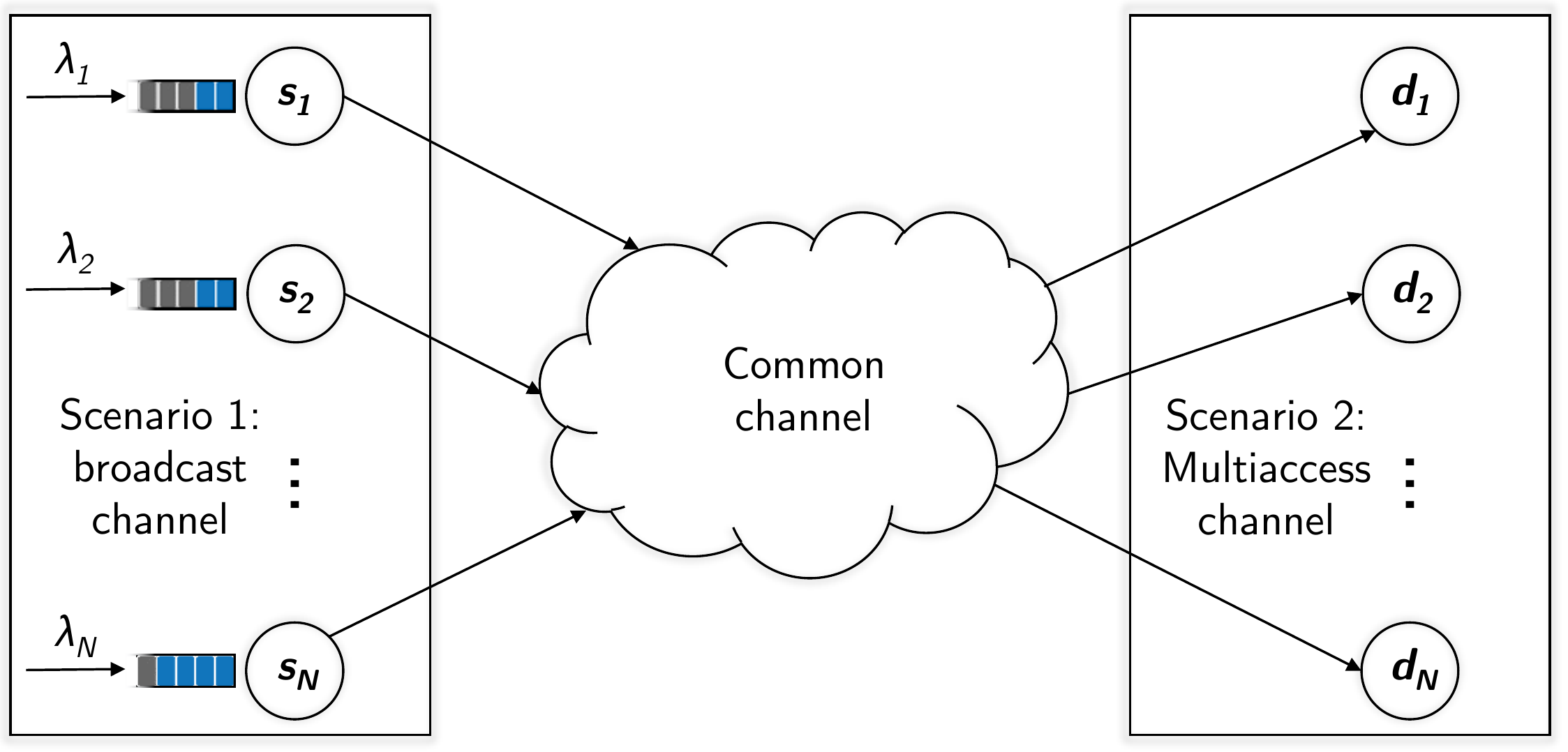}
\caption{A general status update system architecture with $N$ source-destination pairs sharing one common channel.}
\label{fig_arch}
\end{figure}
There have been extensive efforts in the literature for AoI optimization in various scenarios. Considering a single source-destination pair, the queuing theory has been applied by many works to analyze and optimize the AoI performance (cf. \cite{kaul12,huang15,najm17}); the problem of scheduling multiple source-destination pairs has been mainly formulated and addressed with Markov decision process (MDP) and restless multi-armed bandit (RMAB) under different channel models and system assumptions \cite{kadota16,hsu18,jiang18_iot}; a considerable amount of efforts have also been dedicated to, e.g., considering energy harvesting sources, general service processes ($M/G/1$) and so forth (cf. a compressive and timely paper collection on \cite{sun_url}). 

In this paper, firstly towards obtaining the optimal performance, we consider a general status update scheduling problem where multiple source-destination pairs with randomly generated status packets share a common channel which is depicted in Fig. \ref{fig_arch}; note that both wireless broadcast channels and multiaccess channels can be incorporated in this model by considering co-located source nodes and destination nodes, respectively. Our goal in this part is to find the optimal (centralized) status update policy, including the scheduling decision and packet management policy, which minimizes the infinite-horizon time-average AoI. Towards this end, we derive the Whittle's index \cite{whitt84} of the problem in closed-form and establish the corresponding indexability. Based on Weber and Weiss \cite{web90} and numerous evaluation results in the literature, the Whittle's index policy is considered near-optimal, especially when the number of terminals is large. Furthermore, we then focus on the multiple-access channel which is of key interests in future IoT systems. A decentralized status update scheme, i.e., the index-prioritized random access policy (IPRA), is proposed based on the derived Whittle's index and shows universally near-optimal performance. 

    %The rest of the paper is organized as follows. In Section \ref{sec_sm}, we describe the system model and status update design problem in details. In Section \ref{sec_whittle}, we derive the index policy; specifically, we first formulate the MDP problem and then derive the Whittle's index and its indexability. In Section \ref{sec}Section \ref{sec_nr} presents numerical simulation results. Finally, Section \ref{sec_con} concludes the work.

\subsection{Related Work}
The decentralized scheduling problem for AoI optimization is considered by Jiang \emph{et al.} \cite{jiang18_iot}, where a round-robin policy with one-packet buffers policy (RR-ONE) is shown to achieve asymptotically optimal performance with closed-form achievable AoI expressions. However, in the non-asymptotic regime, RR-ONE exhibits notable performance degradation whereas our proposed IPRA is universally near-optimal.

The Whittle's index has been adopted to solve the \emph{centralized} scheduling problem for AoI optimization \cite{kadota16,hsu18}. Kadota \emph{et al.} \cite{kadota16} address the deterministic packet arrival scenario comprehensively where they derive the index policy and also prove its performance bound. We adopt similar methodologies and generalize to stochastic packet arrival scenarios. In parallel with this work, we find that Hsu \cite{hsu18} also derives the Whittle's index with random packet arrivals; however, a specific no-buffer packet management policy is considered whereby the age of a packet at terminals either does not exist  (no arrival at current time slot) or equals one (a packet just arrived). Our work generalizes the work \cite{hsu18} to allow arbitrary packet buffering policy and shows evident performance gain by using only one-packet buffers. Moreover, the index derivation in this paper is more challenging due to the fact that a two-dimensional system state is involved (the states in \cite{hsu18} are considered ``nearly'' one-dimensional since the age of packets is either zero or one). Additionally, it is shown that the index expression in \cite[Theorem 7]{hsu18} coincides with a special case of our results.

\section{System Model and Problem Formulation}
\label{sec_sm}
We consider a general scenario where one central controller is either collecting or broadcasting status update packets for multiple terminals. There are a total of $N$ terminals, e.g., sensors and actuators. A time slotted system is considered. The status updates are conveyed by randomly generated packets at each terminal, reflecting the current status information sensed by terminals and stored at terminal queues. The packet arrivals are modeled by independently identically distributed (i.i.d.) Bernoulli processes with mean rates $\lambda_n\in [0,1]$, $\forall n=\{1,...,N\}$; unlike conventional systems, these packets are usually short, containing a small amount of information however requiring very stringent timeliness. Therefore, instead of throughput, we adopt the AoI metric as our main optimization target. Concretely, the $\tau$-horizon time-average AoI of the system is defined by
\begin{equation}
\label{AoI}
    \Delta_{\pi}^{(\tau,N)} \triangleq \frac{1}{\tau N}\sum_{t=1}^\tau \sum_{n=1}^N \mathbb{E}[h_{n,\pi}(t)],
\end{equation}
where $\pi$ denotes an admissible policy, $\tau$ is the time horizon length, and $h_{n,\pi}(t)$ denotes the AoI of terminal-$n$ at the $t$-th time slot under policy $\pi$. In particular, the long-time-average AoI of the system is concerned, which is defined by
\begin{equation}
\label{aoi_inf}
    \bar{\Delta}_{\pi,N} \triangleq \limsup_{\tau \to \infty} \Delta_{\pi}^{(\tau,N)}.
\end{equation}

\subsection{Status Update Process}
In order to minimize their AoI, the terminals should decide on a transmission scheduling scheme by which they can update the status in a timely fashion and also avoid collisions. In principle, only one terminal should be scheduled for collision avoidance; however, considering the fact that decisions are made autonomously and terminals cannot coordinate perfectly based on global state information, we include concurrent transmissions in our model and hence possible collisions. The status update decisions include:
\begin{itemize}
    \item 
    \emph{Scheduled terminal set and transmission probabilities}: Decide the set of terminals that is scheduled and the transmission probability $p_i$ of each terminal. 
    \item
    \emph{Packet management}: Once a terminal is scheduled, a status update packet is then transmitted based on a packet management policy; the terminal can choose which packet in its buffer to transmit.  
    %Once a terminal decides to transmit, i.e., it is scheduled and the outcome of the random access with probability $p_i$ is positive, a status update packet (if its queue is not empty) is then %transmitted based on a packet management scheme; the terminal stays silent if it has nothing to update.  
\end{itemize}

Due to the decoupled packet management and terminal scheduling, it is obvious that the optimal packet management scheme is to transmit the most up-to-date packet; this is equivalent to maintaining a one-packet buffer at each terminal and only retains the newest packet. Note that the one-buffer packet management is not necessarily optimal when considering service interruption which, however, does not exist in this work \cite{najm16}.

The evolution of the AoI of terminal-$n$ can be written as
\begin{iarray}
\label{evo}
&& h_{n,\pi}(t+1) =  h_{n,\pi}(t) + 1 - u_{n,\pi}(t) \prod_{j \neq n} (1-u_{j,\pi}(t)) g_{n,\pi}(t), \nonumber\\
\end{iarray}
where $u_{n,\pi}(t)=1$ denotes the terminal-$n$ transmits in this time slot and zero otherwise, and the AoI reduction is denoted by $g_{n,\pi}(t)$ which equals the time duration (time slots) between the generation of the last received packet from terminal-$n$ and the updated packet's generation time, i.e.,
\begin{equation}
    g_{n,\pi}(t) = h_{n,\pi}(t) - a_{n,\pi}(t),    	
\end{equation}
where $a_{n,\pi}(t)$ denotes the age of the packet at terminal-$n$ in the $t$-th time slot. Note that $g_{n,\pi}(t)$ equals zero if terminal-$n$ has no packet to update; to unify the notation, we prescribe $a_{n,\pi}(t) = h_{n,\pi}(t)$ in this circumstance and hence $g_{n,\pi}(t) \ge 0$. 

\begin{figure}[!h]
\centering
\includegraphics[width=0.48\textwidth]{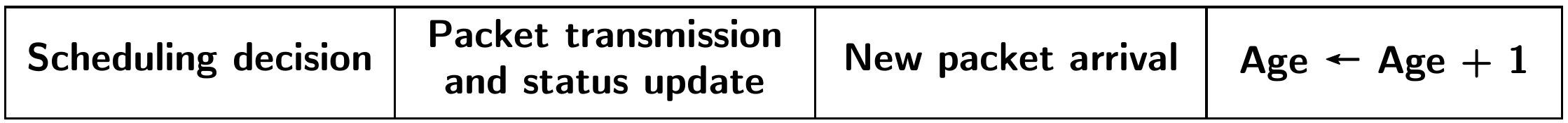}
\caption{Status update sequence. }
\label{fig_sq}
\end{figure}

To be clear, the sequence of events is illustrated in Fig. \ref{fig_sq}.
The AoI (the cost in the MDP formulation introduced later) at time $t$ is defined as the AoI at the time between status update and new packet arrival, which is in line with the post-action age definition in \cite{hsu18}. The age of the newly arrived packet is zero, followed by immediate increment at the end of the sequence, and hence the age of a new packet is one at the time of decision. 

\section{Decoupled Model and Index Policy}
\label{sec_whittle}
In this section, we first formulate the general status update scheduling problem into a MDP problem which is solved by relative value iterations to give a performance benchmark in the simulation section. Afterwards, we introduce the decoupled model where each terminal is examined separately and develop the Whittle's index policy. 
\subsection{MDP-Based Problem Formulation}
The system state is denoted by
\begin{equation}
    \mathcal{S}(t) = \{(a_1(t),d_1(t)),...,(a_N(t),d_N(t))\},
\end{equation}
where $d_n(t)$ denotes the difference between the AoI of terminal-$n$ and the age of the packet of terminal $n$, i.e.,
\begin{equation}
    d_n(t) = h_{n}(t) - a_n(t).
\end{equation}
Note that this state definition is equivalent with the one with $(a_n(t),h_n(t))$, whereas much more convenient in the following derivations and therefore adopted. Based on the fact that the AoI is the age of the packet of the last-updated packet's age, it is clear that $d_n(t) \ge 0$, $\forall n,t$; in addition, we define a new packet's age as one and hence $a_n(t) \ge 1$, $\forall n,t$. The action space is 
\begin{equation}
    \mathcal{U} =  \{1,...,N\},
\end{equation}
which denotes the scheduled terminal index. Note that with global information, it is without loss of optimality to only consider work-conserving non-collision policies \cite[Definition 1]{jiang18_iot}. The state transition probability for terminal-$n$ when not scheduled is described by
\begin{iarray}
&& \Pr\{(a_n,d_n) \to (a_n+1,d_n) \} = 1-\lambda_n; \nonumber\\
&& \Pr\{(a_n,d_n) \to (1,d_n + a_n) \} = \lambda_n,
\end{iarray}
and when scheduled,
\begin{iarray}
&& \Pr\{(a_n,d_n) \to (a_n+1,0) \} = 1-\lambda_n; \nonumber\\
&& \Pr\{(a_n,d_n) \to (1,a_n) \} = \lambda_n,
\end{iarray}
with $a_n \ge 1$ and $d_n \ge 0$. The objective is to minimize the long-time average AoI, i.e., 
\begin{equation}
    \min_{\pi \in \mathcal{U}} \limsup_{\tau \to \infty} \frac{1}{\tau N}\sum_{t=1}^\tau \sum_{n=1}^N \mathbb{E}\left[a_{n,\pi}(t) + d_{n,\pi}(t)(1-u_{n,\pi}(t))\right].
\end{equation}
Such an MDP problem can be solved by the relative value (policy) iteration method with average cost function \cite{bersk}, however, the curse of dimensionality and lack of insights limit the effectiveness of the solution. To address this issue, we note that the above problem can be essentially viewed as an RMAB problem whereby each arm represents one terminal and the reward of pulling an arm is the AoI reduction of the corresponding terminal. It is well-known that the Whittle's index policy is near-optimal for RMAB problems with a large number of arms, in this case terminals \cite{web90}. Therefore, we seek for the index policy in the following subsection. 
\subsection{Decoupled Model}
To design the index policy, a decoupled model is formulated where the AoI of each terminal is compared with an arm with constant cost $m$. Whittle's methodology is that by comparing with an arm with constant cost, the value of each arm is represented by the minimum cost that makes the pulling decision of the arm equally beneficial. Therefore, each arm can be investigated separately, and hence the complexity of finding a solution decreases from exponential with $N$ (MDP value iteration) to linear with $N$. Mathematically, the Whittle's index is equivalent to solving a relaxed version of the Lagrange dual problem \cite{whitt84}. However, the main challenge is that the Whittle's index policy is only defined for problems that are indexable, meaning that the value of each state of an arm can be fully characterized by the constant cost of index policy; on the other hand, the existence of such indexability is problem-dependent and usually difficult to establish, especially with multi-dimensional system states whereby simple structures of the solution (e.g., threshold-based) do not exist.

Since only one terminal is considered, we omit the terminal index in this subsection. Concretely, the decoupled problem is formulated by adding a constant cost $m$ whenever the terminal is scheduled, the objective function of the decoupled model is therefore
\begin{equation}
    \hat{J}^* = \min_{u(t) \in \{0,1\}} \frac{1}{\tau}\sum_{t=1}^\tau \mathbb{E}[a(t) + d(t)(1-u(t)) + m u(t)].
\end{equation}
We consider the long-time average where $\tau \to \infty$. The action is binary, i.e., $u(t)=0$ or $u(t)=1$. This MDP is an average cost problem with infinite horizon and countably infinite state space, and hence the existence of a deterministic and stationary optimal policy is problem-dependent \cite{bersk}. However, we note that this specific problem can be shown to have an optimal deterministic and stationary policy by checking that the assumptions outlined in \cite[Theorem]{lin89} are met. 

Therefore, the steady states exist and the optimal policy can be obtained by solving the Bellman equations to minimize the long-time average AoI. The Bellman equations are given by (since steady states are considered, the time index is omitted)
\begin{iarray}
\label{c2go}
&& f(a,d) + \hat{J}^*  = \nonumber\\
&& \min \left\{ \begin{array}{l}
d + a + (1 - \lambda )f(a + 1,d) + \lambda f(1,d + a),\\
a + m + (1 - \lambda )f(a + 1,0) + \lambda f(1,a)
\end{array} \right\}, 
\end{iarray}
where the top equation in the minimization corresponds to idle and the bottom denotes scheduled, the optimal average cost is denoted by $\hat{J}^*$, and $f(a,d)$ is the differential cost-to-go function and we prescribe $f(1,0)=0$. In what follows, we will solve the above Bellman equations. 

\begin{theorem}
\label{thm1}
Considering the decoupled model, given an auxiliary cost $m$, the optimal action with a state $(a,d)$ is to schedule the terminal when $d \ge D_a$ and idle otherwise. Specifically, $D_1=\beta$ where $\beta$ is the unique positive solution to the following equation:
\begin{iarray}
    \frac{1}{2}\beta^2 + \left(\frac{1}{\lambda}-\frac{1}{2}\right)\beta-m = 0. \nonumber\\
\end{iarray}
The rest of the thresholds are given by
\begin{iarray}
    D_a = \left\{\,
        \begin{IEEEeqnarraybox}[][c]{l?s}
        \IEEEstrut	
        (1-\lambda+a\lambda) \beta -\lambda \frac{(a-1)a}{2}, &  if $1 \le a < \beta$;\\
    	\lambda m, & if $a \ge \beta$.
    	\IEEEstrut
    	\end{IEEEeqnarraybox}
    	\right.
\end{iarray}
The optimum average AoI is 
\begin{iarray}
\hat{J}^* = \frac{1}{\lambda}+\beta. \quad\square
\end{iarray}
\end{theorem}
\begin{IEEEproof}
The basic methodology to solve the equations is to first assume the structure of the solution and the optimal policy, and then solve the Bellman equations based on the assumptions. Consistency of the solutions and the assumptions should be checked afterwards. Since there is (with mild conditions) a unique solution to the Bellman equations, such a constructive method can work to find the solution. See Appendix \ref{app_thm1} for details.
\end{IEEEproof}
\begin{remark}
Examining the difference between Theorem \ref{thm1} and \cite[Theorem 5]{hsu18}, due to the no-buffer assumption in \cite{hsu18}, the system states are nearly one-dimensional (age of a packet is either one or none) and hence there is only one AoI threshold; in contrast, there is one threshold for \emph{each} $a$ (age of the packet at terminal-side) in Theorem \ref{thm1}, making the derivation of the theorem considerably more challenging. On the other hand, we will show based on simulation results that by using only one-packet buffers, the performance gain is evident compared with the no-buffer index policy.$\hfill\square$
\end{remark}

The indexability of the index policy can be readily derived based on Theorem \ref{thm1}.
\begin{defi}[Indexability]
Given costs $m_1$ and $m_2$, and the sets of states that the optimal action is to idle are denoted by $\Pi_{m_1}$ and $\Pi_{m_2}$ respectively, the problem is indexable if 
\begin{equation}
    \forall m_1, m_2\textrm{ with } m_1 < m_2 \Rightarrow \Pi_{m_1} \subseteq  \Pi_{m_2},
\end{equation}
and for $m=0$, $\Pi_{m} =  \emptyset$; for $m \to \infty$, $\Pi_{m}$ is the entire state space.
\end{defi}
\begin{theorem}[Indexability]
\label{thm2}
Consider the decoupled model and the scheduling policy $\pi_{\mathsf{D}}$ given in Theorem \ref{thm1}, then $\pi_{\mathsf{D}}$ is indexable. $\hfill\square$
\end{theorem}
\begin{IEEEproof}
See Appendix \ref{app_thm2} for details.
\end{IEEEproof}

The index for any state is described as follows.
\begin{theorem}[Whittle's Index]
\label{def_index}
Consider the decoupled model and denote the index by $m(a,d)$ with state $(a,d)$, 
\begin{iarray}
\label{index_e}
m(a,d) = \left\{\,
        \begin{IEEEeqnarraybox}[][c]{l?l}
        \IEEEstrut	
        \frac{1}{2} x^2 + \left(\frac{1}{\lambda}-\frac{1}{2}\right) x, &\textrm{if } d > \frac{\lambda}{2}a^2+\left(1-\frac{\lambda}{2}\right)a; \\
    	\frac{d}{\lambda},&\textrm{otherwise}, 
    	\IEEEstrut
    	\end{IEEEeqnarraybox}
    	\right. \nonumber\\
\end{iarray}
where 
\begin{iarray}
x\triangleq \frac{d + \frac{a(a-1)}{2}\lambda}{1-\lambda+a\lambda}. \quad\square
\end{iarray}
\end{theorem}
\begin{remark}
The derivation of the index follows directly from the optimal policy of the decoupled model in Theorem \ref{thm1}, by the reasoning that the index of a state equals the minimum auxiliary cost that makes the scheduling actions of the terminal under the current state equally beneficial.$\hfill\square$
\end{remark}
\begin{remark}
The index is a generalization of the previous results in \cite{kadota16} where the index without randomly generated status packets is derived. In their work, the index is (with transmission success probability $p=1$, user weight $\alpha=1$ and frame length $T=1$ in \cite{kadota16})
\begin{iarray}
\label{h1}
C(h) = \frac{1}{2}h(h+1),
\end{iarray}
where $h$ is the AoI of the terminal. Based on \eqref{index_e}, with $\lambda=1$ and hence the packet age is $a=1$, we obtain
\begin{iarray}
\label{h2}
m(1,d) &=& \frac{1}{2}d^2 + \left(\frac{1}{\lambda}-\frac{1}{2}\right)d = \frac{1}{2} d(d+1).
\end{iarray}
The difference between $h$ in \eqref{h1} and $d=h-1$ in \eqref{h2} is due to the fact that \cite{kadota16} adopts the pre-action age and we adopt the post-action age; either case does not affect the results much. The derived index also coincides with \cite{hsu18} when $a=1$. Therefore, it is observed that our results are consistent with previous work and generalize to the scenario with random packet arrivals and arbitrary buffering strategy. $\hfill\square$
\end{remark}

\section{Index-Prioritized Random Access Policy}
\label{sec_ipra}
The index policy derived above clearly requires global information (all the age and AoI information) for scheduling decisions, and hence it is recognized as a centralized scheduling policy which makes it undesirable in wireless uplinks due to signaling overhead concerns. Nonetheless, inspired by the index policy, we describe an index-prioritized random access policy (IPRA) which achieves the same universally near-optimal AoI performance as the index policy, however, with decentralized protocol structure. A key observation is that each terminal can calculate its own index, denoted by $I_n$, based on its transmission history and packet arrivals. Thereby, this individual index is mapped to a transmission probability based on a public mapping function which captures the idea that only valuable packets (packets with high index value) are transmitted; a random access (contention) period is hence introduced to resolve possible collisions. The selection of the public mapping function is tricky and we propose to use a single-threshold function which, notwithstanding its simplicity, achieves near-optimal performance based on simulation results. The detail procedure is described in Algorithm \ref{alg:ipra}.

\begin{algorithm}[!h]
	\caption{IPRA}
	\label{alg:ipra}
		\textbf{{Contention period: For $n \in \{1,...,N\}$}}\\
		\If{$I_n \ge \mathsf{indexThreshold}$}{Terminal-$n$ transmits with probability $p$.}
		\Else{Terminal-$n$ is idle.}
		\textbf{Transmission/Collision frame:}\\
		If the transmission is successful, the central controller feeds back an ACK; otherwise a NACK is fed back.\\
		Go to the contention period.		
\end{algorithm}
\begin{figure}[!h]
\centering
\includegraphics[width=0.48\textwidth]{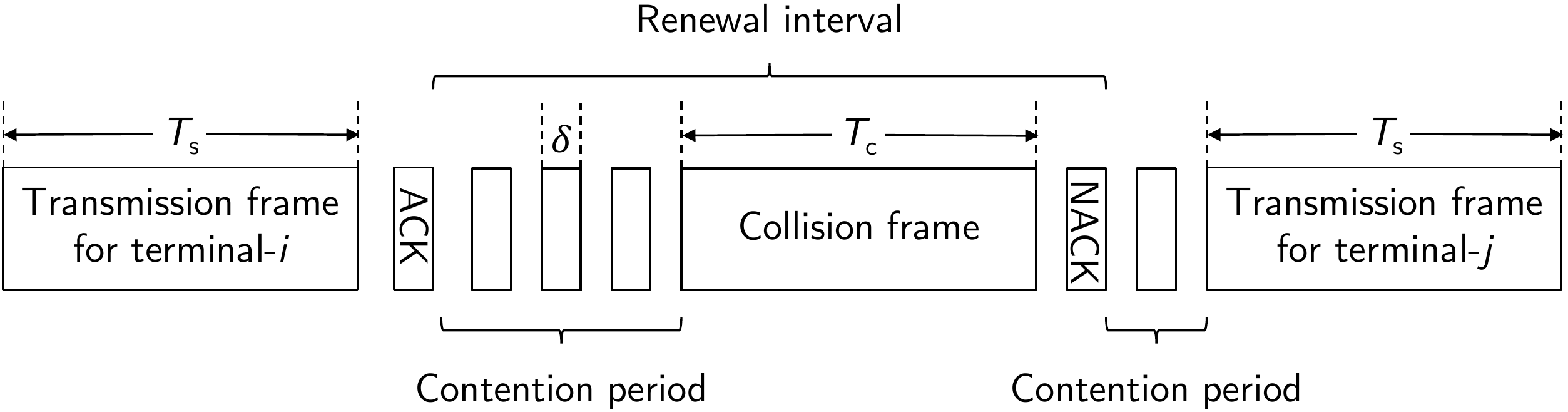}
\caption{Transmission timeline of IPRA.}
\label{fig_ipra}
\end{figure}

In IPRA, two parameters, i.e., transmission probability $p$ and index threshold $\mathsf{indexThreshold}$, need to be optimized. In our implementation we find that the optimization is usually easy since the performance as a function of these parameters, i.e., $f(p,\mathsf{indexThreshold})$, is quasi-concave (unimodal). Therefore, well-known methods such as \cite{comb14} can be applied.

Furthermore, we will show in the following that the overhead of IPRA, which includes the contention period and collision frames, is negligible. Denote the transmission frame length and collision frame length as $T_\mathsf{s}$ and $T_\mathsf{c}$ respectively, and $\delta$ as the length of one contention slot. We assume $T_\mathsf{s} = T_\mathsf{c}$ which is the worst-case assumption meaning that a terminal can only finds out the transmission fails after one entire transmission frame. 

\begin{theorem}
\label{thm3}
\begin{equation}
    \lim_{K \to \infty} \frac{\sum_{k=1}^K T_\mathsf{s} \mathds{1}_{\mathsf{s},k} }{\sum_{k=1}^K \tau_k} \to 1,\textrm{ with } \frac{T_\mathsf{s}}{\delta} \to \infty,
\end{equation}
where $\tau_k$ denotes the length of the $k$-th renewal interval, and $\mathds{1}_{\mathsf{s},k}$ is a indicator function denoting whether the transmission is successful in the $k$-th renewal interval. $\hfill\square$
\end{theorem}
\begin{IEEEproof}
The proof is based on the elementary renewal theory and previous work on CSMA in, e.g., \cite{gai11}. The details are omitted due to lack of space. Moreover, the ratio of $\frac{T_\mathsf{s}}{\delta}$ is usually very high (around $100$ is common \cite{gai11}) and hence the overhead of IPRA is negligible.
\end{IEEEproof}
\section{Simulation Results}
\label{sec_nr}
In this section, we present numerical results based on computer simulations which run the scheduling policies for $10^6$ time slots and obtain the time-average AoI. It is no surprise that the Whittle's index policy is near-optimal given many existing works which have validated this conclusion in various scenarios. In view of this, the main purpose of the section is to highlight the performance difference between the Whittle's index policy with optimal packet management (one-packet buffers) and several existing scheduling policies \cite{jiang18_iot,hsu18}, and to demonstrate the performance of IPRA. 

\begin{figure}[!t]
    \centering    
    \subfigure{\includegraphics[width=0.225\textwidth]{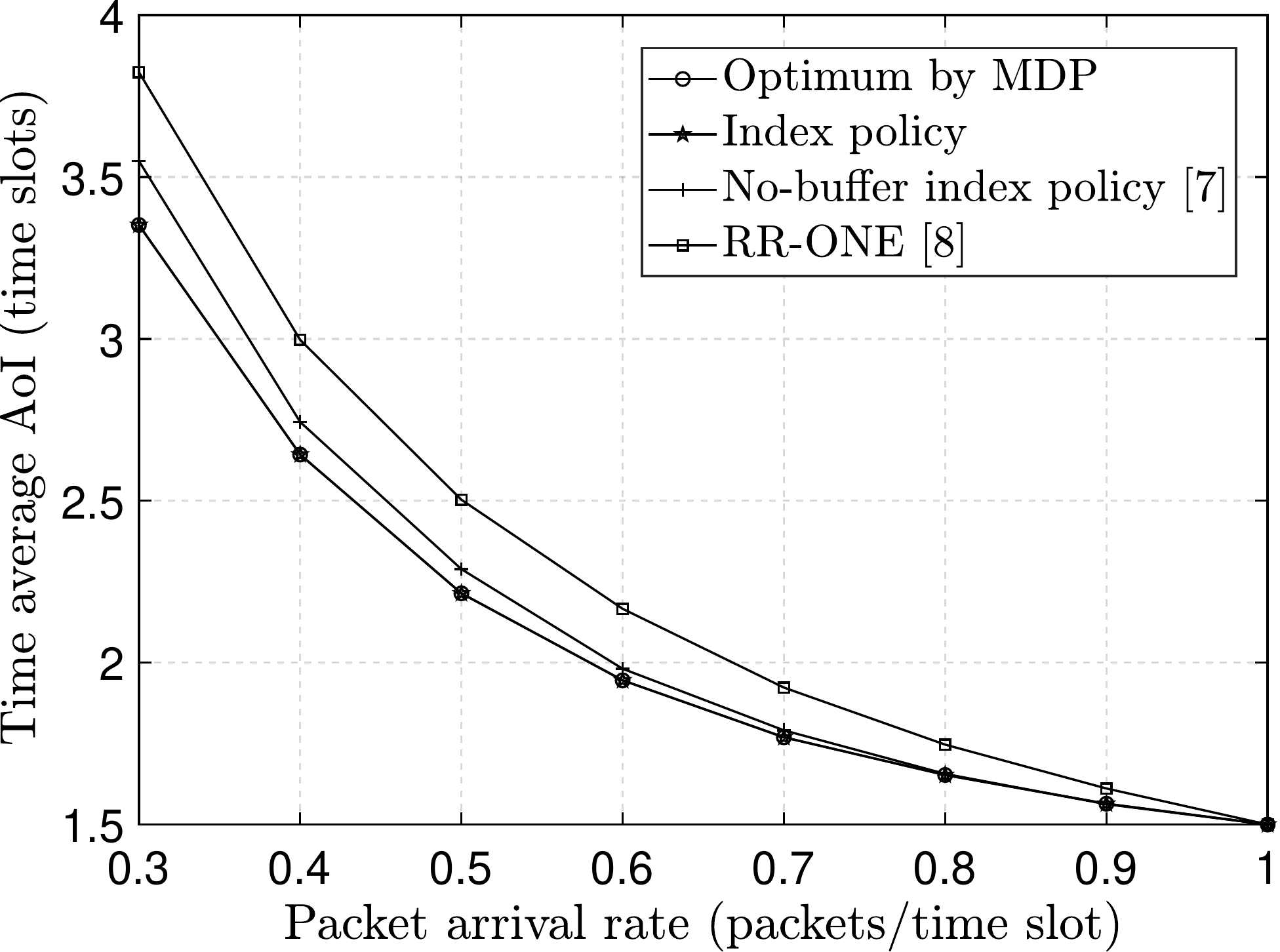}}
    \subfigure{\includegraphics[width=0.225\textwidth]{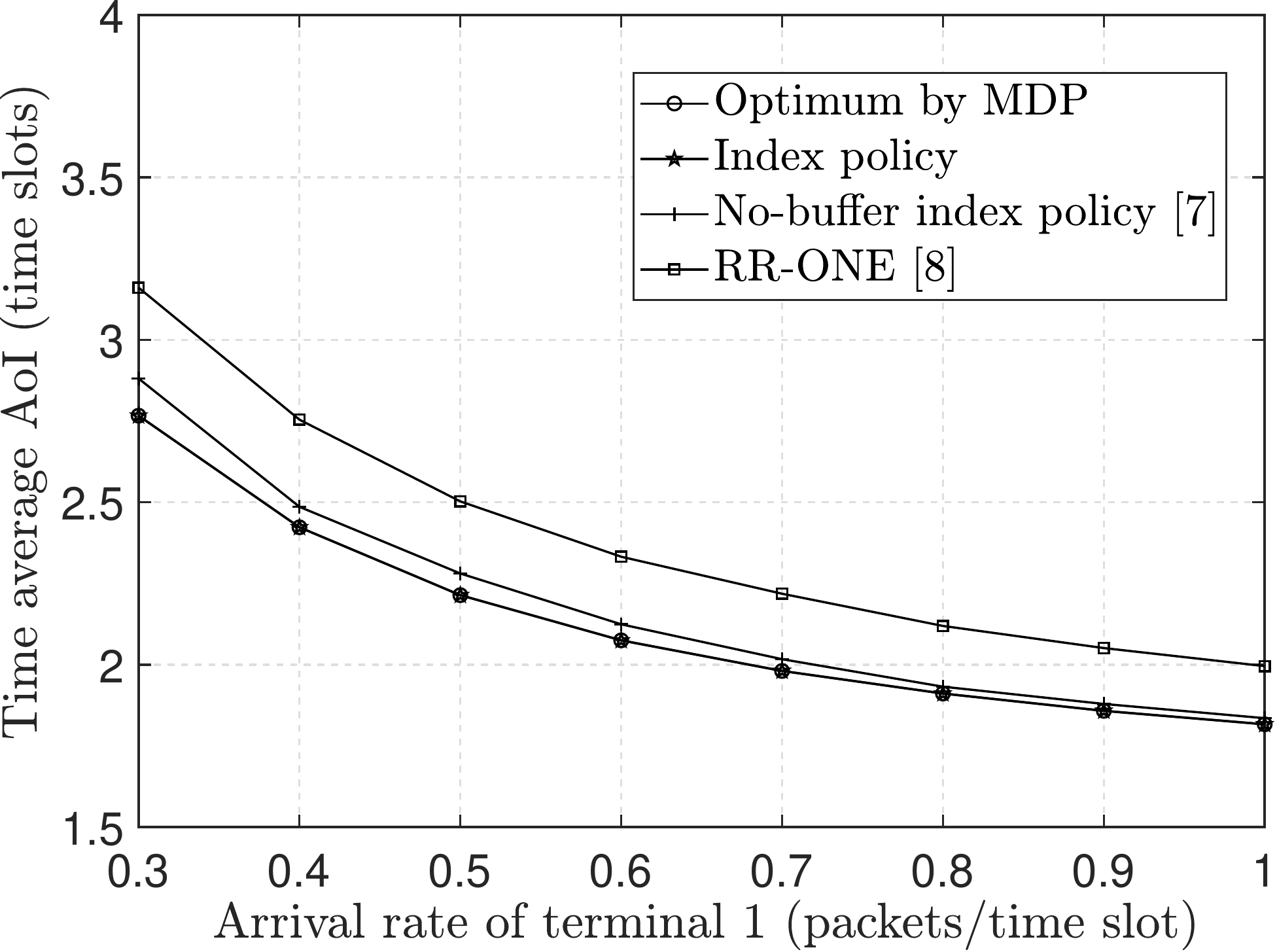}}
    \caption{Performance comparisons with optimum by solving the MDP and $2$ terminals having identical packet arrival rate (left) and heterogeneous arrival rates (right) with terminal $1$'s arrival rate shown as x-axis and terminal $2$'s fixed as $0.5$.}
    \label{fig_mdp}
\end{figure}

First in Fig. \ref{fig_mdp}, a $2$-terminal case is considered and the optimum AoI based on solving the MDP via relative value iteration is obtained. The difference between the Whittle's index policy and the optimum is hardly visible, whereas the no-buffer index policy suffers evident performance loss when the packet arrival rate is low. This can be explained that the no-buffer strategy drops precious (when with low arrival rates) status packets when both terminals have arrivals in the same time slot; in this case packets should be stored for update in the future. Note that the no-buffer index policy is identical with the index policy in this paper when $\lambda$ approaches one, by showing that the index in Theorem \ref{thm2} in this regime, i.e., \eqref{h2}, coincides with the index in \cite{hsu18}.\footnote{The difference between $h$ here and $h+1$ in \cite{hsu18} is due to the difference in the definitions of initial packet age.}

\begin{figure}[!t]
\centering
\includegraphics[width=0.45\textwidth]{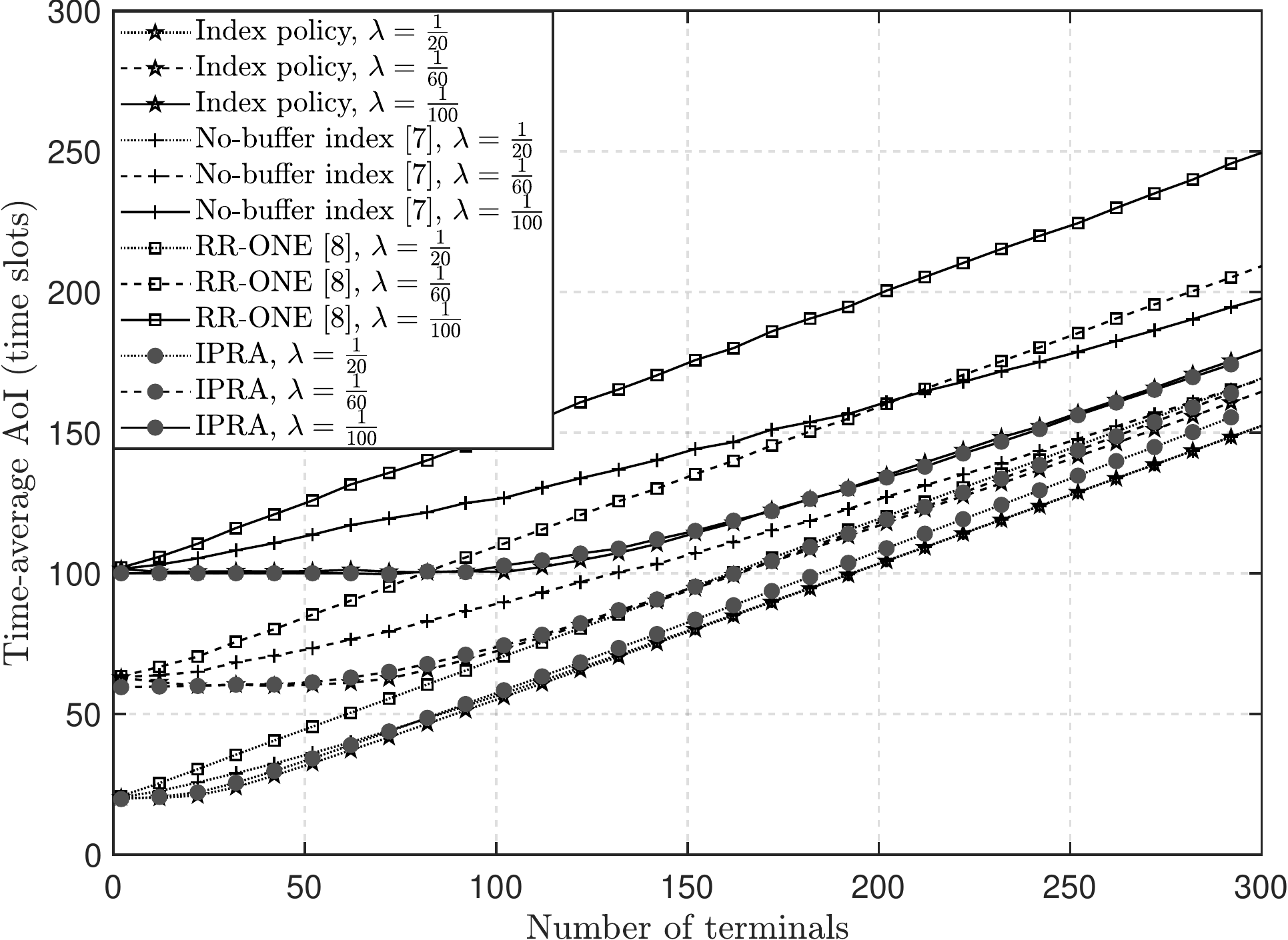}
\caption{Performance comparisons with a large number of terminals with identical packet arrival rate of $\lambda$ specified in the legend.}
\label{fig_indN}
\end{figure}

The performance gap between one-buffer and no-buffer index policies is more pronounced in Fig. \ref{fig_indN}, where we also simulate the decentralized policy RR-ONE in \cite{jiang18_iot} and IPRA. It has been shown in \cite{jiang18_iot} that RR-ONE achieves the optimal scaling factor with a large $N$ but is suboptimal with finite $N$; this is observed in the figure. It is shown that the gap between no-buffer and one-buffer policies is most evident when the mean arrival interval is comparable with the number of terminals, i.e., the joint-asymptotic regime where $N \to \infty,\,\frac1\lambda \to \infty$, and $N\lambda \to C$, where $C$ is a fixed constant. This is because in this regime, the delays due to scheduling among terminals and random packet arrivals are equally significant and hence neither can be ignored, making the optimal policy most elusive; when $N \to \infty$ the delay due to random packet arrivals can be neglected and hence dropping packets by no-buffer policy is also near-optimal.

More importantly, we find in Fig. \ref{fig_indN} that IPRA achieves universally near-optimal performance by noting that its performance is very close to the one-buffer index policy. The length of one contention slot is set to be $1/100$ \cite{gai11} of a transmission frame which is of the same length of a time slot in RR-ONE and index policies. The transmission probability and index threshold in IPRA are optimized based on a bi-section search algorithm.

\begin{figure}[!t]
\centering
\includegraphics[width=0.45\textwidth]{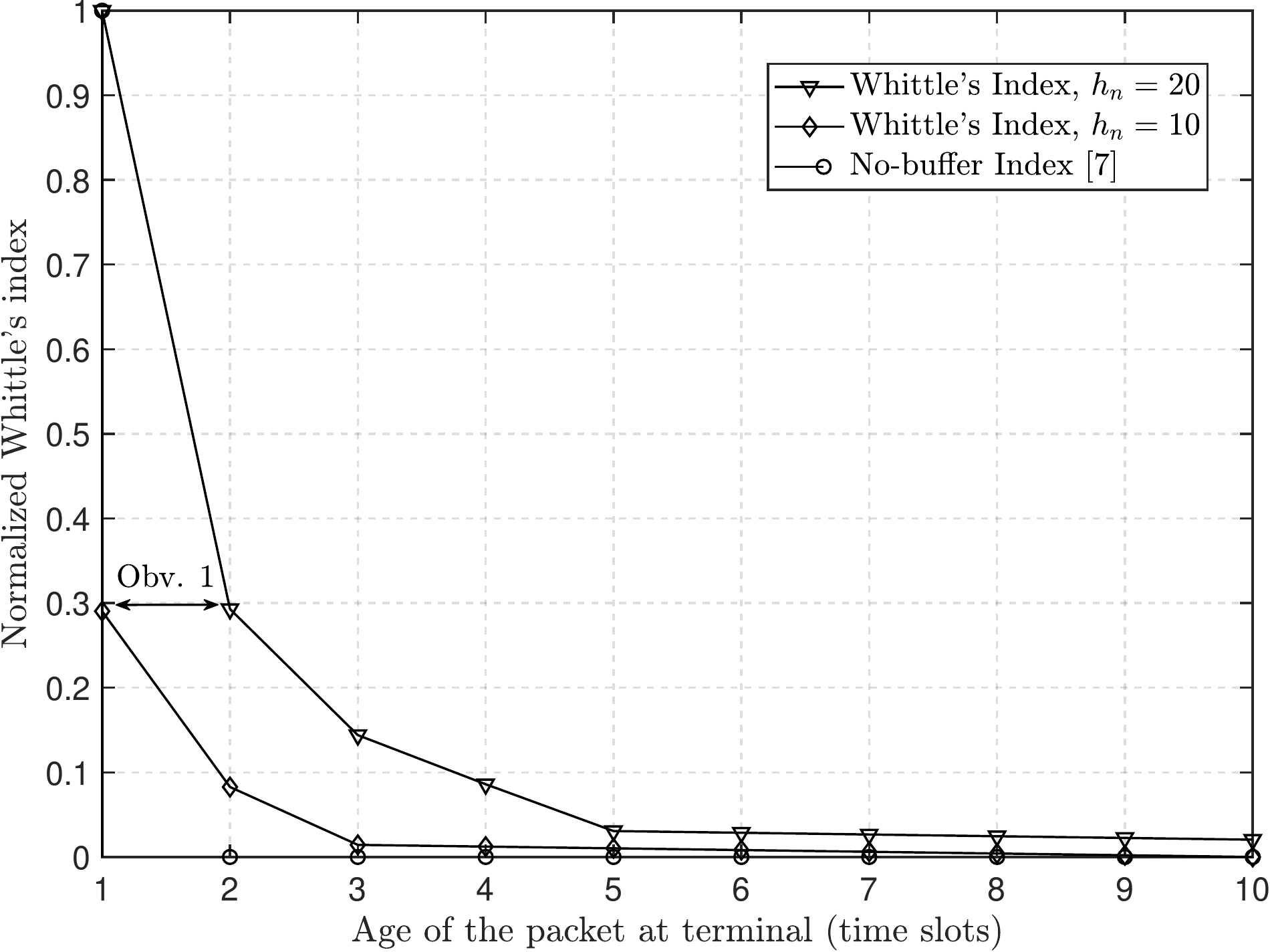}
\caption{Illustration of the difference between the Whittle's indices with optimal buffering strategy and no-buffer.}
\label{fig_indC}
\end{figure}

To demonstrate the benefit of buffering packets over no-buffer policies, we compare the normalized indices with various $a$; note that only the relative value of the index is relevant hence justifying the normalization. The benefit of buffering is shown by the fact that the index value of a state with the packet's age larger than one is still significant; hence dropping the packet of such kind losses performance. In particular, based on observation $1$ in the figure, a state with $(a,h)=(1,10)$ has approximately the same index with a state $(2,20)$, meaning that scheduling a terminal with state $(2,20)$ (buffering the packet for one time slot) is equally beneficial with one with $(1,10)$, whereas the no-buffer index deems the state $(2,20)$ as zero-value which is clearly sub-optimal.

\section{Conclusions}
\label{sec_con}
In summary, the answer to the question whether decentralized status update can achieve universally near-optimal AoI in wireless multiaccess channels is positive based on the following presented results: (a) The centralized Whittle's index policy which has near-optimal performance is derived in closed-form and the indexability is established. The performance thereof serves as a performance benchmark for optimality. (b) IPRA adopts the derived index to prioritize packet transmissions among terminals and a random access procedure is applied with proven negligible overhead, such that the status update is decentralized. Moreover, it is shown that the performance of IPRA is nearly identical with the index policy and outperforms state-of-the-arts in the literature.

For future IoT applications where the main objective is to optimize the status update timeliness in wireless uplinks, IPRA is a promising decentralized multiple-access solution achieving better information freshness with smaller signaling overhead compared with current approaches, e.g., CSMA and grant-based schemes in cellular systems. Future work includes considering arbitrary packet arrival patterns such as bursty arrivals, as well as non-Markovian information sources.

\section*{Acknowledgement}
This work is sponsored in part by the Nature Science Foundation of China (No. 61701275, No. 91638204, No. 61571265, No. 61621091), the China Postdoctoral Science Foundation, and Hitachi Ltd.

\appendices
\section{Proof for Theorem \ref{thm1}}
\label{app_thm1}
As a first step, given an arbitrary auxiliary cost $m \ge 0$, two assumptions regarding the optimal policy and the solution are made.
\begin{assump}
\label{ass1}
The optimal policy is a threshold policy, denoted by $\pi_{\mathsf{D}}$, that is given any state $(a,d)$, the optimal action is to idle when $0 \le d < D_a$ and to schedule when $d \ge D_a$. Furthermore, the thresholds satisfy
\begin{equation}
\label{th}
    D_1 \le D_2 \le...\le D_a \le... \quad\hfill\square
\end{equation}
\end{assump}
\begin{assump}
\label{ass2}
For any $a$,
\begin{equation}
\label{mon}
    f(a,0) \le f(a,1) \le...\le f(a,d) \le... \quad\hfill\square
\end{equation}
\end{assump}

Both assumptions are in fact intuitive as explained in the following. The threshold-based policy structure in Assumption \ref{ass1} stems from the tradeoff between auxiliary cost and AoI reduction ($d$) in each time slot; the monotonicity of the thresholds in \eqref{th} reflects the fact that with a relatively older packet (larger $a$) at terminal-side, it is better wait for a new packet arrival than to schedule this time. The Assumption \ref{ass2} simply states the fact that a state with a larger AoI and the same packet age has a larger cost-to-go value. Although reasonable, after we solve the Bellman equations, we will check the consistency of these assumptions. For ease of exposition, we assume the thresholds can be real values in this section.

With $\pi_{\mathsf{D}}$, an important property regarding the differential cost-to-go function is stated below. 
\begin{prop}
\label{prop1}
For any $a_1,a_2\ge1$, $0 \le d_1< D_{a_1}$, $0 \le d_2< D_{a_2}$, and $a_1+d_1=a_2+d_2$, we have 
\begin{equation}
    f(a_1,d_1)=f(a_2,d_2). \quad\hfill\square
\end{equation}
\end{prop}
\begin{IEEEproof}
To prove this proposition, we first prove the following lemma.
\begin{lemma}
\label{lm1}
There exists $a_{\mathsf{M}}$ such that $\forall a \ge a_{\mathsf{M}}$, the threshold $D_a=D_{a_\mathsf{M}}$ is a constant. $\hfill\square$
\end{lemma}
\begin{IEEEproof}
Based on \eqref{th}, the thresholds are monotonically increasing with respect to $a$. Note that when $d \ge m$, observing the cost-to-go equations in \eqref{c2go}, we can show that the optimal action in this case is to schedule the terminal, that is the upper part in the minimization (denoted by $\mu_0(a,d)$) is larger than the lower part (denoted by $\mu_1(a,d)$). Concretely, with $d \ge m$,
\begin{iarray}
    && \mu_0(a,d) - \mu_1(a,d) \nonumber\\
    &=& d-m + (1-\lambda)(f(a+1,d)-f(a+1,0)) \nonumber\\
    && + \lambda(f(1,d+a)-f(1,a)) \ge 0.
\end{iarray}
The equality is based on Assumption \ref{ass2}. This means that $\forall a$, $D_a \le m$. Therefore, the threshold array $\{D_a:\,a=1,2,...\}$ is monotonically increasing with a finite upper bound; hence the limitation exists and this concludes the proof.
\end{IEEEproof}

Given $\pi_{\mathsf{D}}$ and based on Lemma \ref{lm1}, the action when $a \ge a_{\mathsf{M}}$ and $d \ge D_{a_\mathsf{M}}$ is to schedule the terminal. Therefore, based on \eqref{c2go} and Assumption \ref{ass1} we have $D_{a_\mathsf{M}} \ge D_1$, and hence
\begin{equation}
\label{1d}
    f(1,d)+ \hat{J}^* = 1+m+(1 - \lambda )f(2,0) + \lambda f(1,1),\, d \ge D_{a_\mathsf{M}}.
\end{equation}
Note that based on \eqref{1d}, $f(1,d) = f(1,D_{a_\mathsf{M}})$, $\forall d \ge D_{a_\mathsf{M}}$. Additionally, the action when $d = D_{a_\mathsf{M}}-1$ and $a \ge a_{\mathsf{M}}$ is to idle, and hence,
\begin{iarray}
    && f(a,D_{a_\mathsf{M}}-1) + \hat{J}^*\nonumber\\
    &=& D_{a_\mathsf{M}}+a-1+ (1 - \lambda )f(a + 1,D_{a_\mathsf{M}}-1) \nonumber\\
    && + \lambda f(1,D_{a_\mathsf{M}} + a-1) \nonumber\\
    &=& D_{a_\mathsf{M}}+a-1+ (1 - \lambda )f(a + 1,D_{a_\mathsf{M}}-1) \nonumber\\
    && +  \lambda f(1,D_{a_\mathsf{M}}).
\end{iarray}
Denote 
\begin{equation}
    g \triangleq D_{a_\mathsf{M}}- \hat{J}^* -1 + \lambda f(1,D_{a_\mathsf{M}}),
\end{equation}
we obtain
\begin{iarray}
\label{rec}
    && f(a,D_{a_\mathsf{M}}-1) = a + (1 - \lambda )f(a + 1,D_{a_\mathsf{M}}-1) +  g.
\end{iarray}
Solving \eqref{rec} recursively with respect to $a$ yields
\begin{iarray}
\label{drope}
    && f(a,D_{a_\mathsf{M}}-1) = \gamma_0{(1 - \lambda )^{ a_{\mathsf{M}} - a}} + \frac{a+g}{\lambda}+\frac{1-\lambda}{\lambda^2}
\end{iarray}
where $\gamma_0 = f(a_{\mathsf{M}},D_{a_\mathsf{M}}-1) - \frac{a_\mathsf{M}+g}{{\lambda}} - \frac{1-\lambda}{\lambda^2}$ and $a \ge a_{\mathsf{M}}$. When $\gamma_0 \neq 0$, the expression has an exponential term which is unreasonable because this means a state with a larger packet age (and the same $d$) has an exponentially larger cost; moreover, we will show the solution with $\gamma_0=0$ is a valid solution by checking the consistency. A condition is used in \cite{kadota18} under similar circumstances by claiming
\begin{equation}
    \lim_{a \to \infty}(1 - \lambda )^{a} f(a,D_{a_\mathsf{M}}-1) =0,
\end{equation}
which essentially carries the same meaning as ours. Likewise, we obtain
\begin{iarray}
\label{ds}
    &&f(a,D_{a_\mathsf{M}}-s) = \frac{a+g-s}{\lambda}+\frac{1}{\lambda^2}, \nonumber\\
    &&a \ge s+a_{\mathsf{M}}-1,\, 1 \le s \le D_{a_\mathsf{M}},
\end{iarray}
Note that, e.g., 
\begin{iarray}
    f(a_{\mathsf{M}},D_{a_\mathsf{M}}-1) &=& f(a_{\mathsf{M}}+1,D_{a_\mathsf{M}}-2) = ... \nonumber\\
    &=& f(D_{a_\mathsf{M}}+a_{\mathsf{M}}-1,0).
\end{iarray}
It is therefore clear that the proposition holds for any $a_1,a_2,d_1,d_2$ satisfying $a_1+d_1=a_2+d_2 \ge D_{a_\mathsf{M}}+a_{\mathsf{M}}-1$. For $a_1,a_2,d_1,d_2$ satisfying $a_1+d_1=a_2+d_2 < D_{a_\mathsf{M}}+a_{\mathsf{M}}-1$, an induction based proof is adopted. 

Suppose the proposition holds for any $a_1,a_2,d_1,d_2$ satisfying $a_1+d_1=a_2+d_2 = D_{a_\mathsf{M}}+a_{\mathsf{M}}-1-s$, $0 \le s \le D_{a_\mathsf{M}}+a_{\mathsf{M}}-2$, $d_1< D_{a_1}$ and $d_2< D_{a_2}$, then for any $a_1^\prime,a_2^\prime,d_1^\prime,d_2^\prime$ satisfying $a_1^\prime+d_1^\prime=a_2^\prime+d_2^\prime = D_{a_\mathsf{M}}+a_{\mathsf{M}}-2-s$ and $d_1^\prime< D_{a_1^\prime}$, $d_2^\prime< D_{a_2^\prime}$, the action is to idle based on $\pi_{\mathsf{D}}$. It follows from \eqref{c2go} that
\begin{iarray}
    &&f(a_1^\prime, d_1^\prime) \nonumber\\
    &=& -\hat{J}^*+d_1^\prime + a_1^\prime + (1 - \lambda )f(a_1^\prime + 1,d_1^\prime) + \lambda f(1,d_1^\prime + a_1^\prime) \nonumber\\
    & \overset{(a)}{=} & -\hat{J}^*+d_2^\prime + a_2^\prime + (1 - \lambda )f(a_2^\prime + 1,d_2^\prime) + \lambda f(1,d_2^\prime + a_2^\prime) \nonumber\\
    &=& f(a_2^\prime, d_2^\prime).
\end{iarray}
The equality $(a)$ is based on the induction hypothesis. Also note that $d_1^\prime< D_{a_1^\prime} \le D_{a_1^\prime+1}$ based on Assumption \ref{ass1}, and hence the conditions are all satisfied. For the induction basis, the proposition holds for $s=0$ based on \eqref{ds}. Therefore, the proposition is concluded.
\end{IEEEproof}

For $1 \le a < D_1$, based on \eqref{c2go},
\begin{iarray}
    f(a,0) &=& -\hat{J}^*+a + (1 - \lambda ) f(a+1,0) + \lambda f(1,a) \nonumber\\
    &=& -\hat{J}^*+a + f(a+1,0),
\end{iarray}
where the last equality is based on Proposition \ref{prop1}. Given that $f(1,0)=0$, it follows that
\begin{equation}
\label{ds3}
    f(a,0) = (a-1)\hat{J}^* - \frac{a(a-1)}{2}, \, 1 \le a \le D_1.
\end{equation}

To proceed, we obtain another important property of the solution in the following proposition. 
\begin{prop}
\label{prop2}
For any state $(a,d)$ with $a \ge 1$ and $d \ge D_a$, 
\begin{equation}
    \label{difm}
    f(a,d) - f(a,0) = m.
\end{equation}
\end{prop}
\begin{IEEEproof}
For some state $(a,d)$ with $a \ge 1$ and $d \ge D_a$, based on \eqref{c2go} and $\pi_{\mathsf{D}}$,
\begin{iarray}
f(a,d) &=& -\hat{J}^* + a +m + (1-\lambda)f(a+1,0) + \lambda f(1,a) \nonumber\\
f(a,0) &=& -\hat{J}^* + a + (1-\lambda)f(a+1,0) + \lambda f(1,a).
\end{iarray}
Therefore the proposition is concluded by observing the difference of the two above equations.
\end{IEEEproof}
\begin{remark}
This proposition is reasonable since for states $(a,d)$ with $d \ge D_a$ and $\pi_{\mathsf(D)}$, the relative cost compared with state $(a,0)$ is a one-time scheduling auxiliary cost $m$. A direct corollary is that 
\begin{iarray}
\label{1dm}
f(1,d) = f(1,0)+m = m,\,d \ge D_1. \quad\square
\end{iarray}
\end{remark}

Based on Proposition \ref{prop1} and the same arguments in \eqref{drope}, we obtain
\begin{equation}
\label{ds2}
    f(a,0) = \frac{a-\hat{J}^*-1}{\lambda}+m+\frac{1}{\lambda^2}, \, a \ge D_1.
\end{equation}
Based on Proposition \ref{prop2}, when $d \ge D_a$ it follows that
\begin{iarray}
\label{2d}
    && f(a,d) \nonumber\\
    &=& m+ f(a,0) \nonumber\\
    &=& \left\{\,
        \begin{IEEEeqnarraybox}[][c]{l?s}
        \IEEEstrut	
        m + a \hat{J}^* - \frac{(a-1)a}{2}, &  if $1 \le a < D_1$;\\
    	\frac{a}{\lambda}+2m-\frac{1-\lambda}{\lambda}\hat{J}^* + \frac{1-\lambda}{\lambda^2}, & if $a \ge D_1$.
    	\IEEEstrut
    	\end{IEEEeqnarraybox}
    	\right.    	
\end{iarray}

Combining \eqref{ds3} and \eqref{ds2} when $d=D_1$ gives us the relationship among $m$, $\hat{J}^*$ and $D_1$, i.e.,
\begin{equation}
    \label{mjd}
    m = \left(D_1-1+\frac{1}{\lambda}\right) \hat{J}^* -\frac{D_1^2}{2} + \frac{D_1}{2} -\frac{D_1}{\lambda} + \frac{\lambda-1}{\lambda^2}.
\end{equation}

Now we have obtained all differential cost-to-go function expressions based on $\pi_{\mathsf{D}}$. To show that this is the solution to the Bellman equations, we need to demonstrate the policy $\pi_{\mathsf{D}}$, whereby we assume the optimal policy has certain properties in Assumption \ref{ass1} and \ref{ass2}, is indeed consistent. In this regard, it will be shown that by carefully arranging the thresholds $D_a$ with $a=1,2,...$, the threshold policy $\pi_{\mathsf{D}}$ is the solution to the Bellman equations in each state. Let us first check the boundary points, i.e., state $(a,D_a)$.

For state $(a,D_a)$, observe the cost-to-go equations in \eqref{c2go}, we need to show that the optimal action in this case is to schedule the terminal, that is the upper part in the minimization (denoted by $\mu_0(a,D_a)$) is larger than the lower part (denoted by $\mu_1(a,D_a)$). First, consider the case where $a \ge D_1$, the difference is
\begin{iarray}
    && \mu_0(a,D_a) - \mu_1(a,D_a) \nonumber\\
    &=& D_a-m + (1-\lambda)(f(a+1,D_a)-f(a+1,0)) \nonumber\\
    && + \lambda(f(1,D_a+a)-f(1,a)) \nonumber\\
    &\overset{(a)}{=}& D_a-m + (1-\lambda)\frac{D_a}{
    \lambda} \ge0,
\end{iarray}
where the equality $(a)$ is attributed to \eqref{1dm} and \eqref{ds2}. Therefore we obtain a condition
\begin{iarray}
\label{cond1}
    D_a \ge \lambda m.
\end{iarray}
The case with $1\le a < D_1$ yields
\begin{iarray}
    && \mu_0(a,D_a) - \mu_1(a,D_a) \nonumber\\
    &=& D_a-m + (1-\lambda)(f(a+1,D_a)-f(a+1,0)) \nonumber\\
    && + \lambda(f(1,D_a+a)-f(1,a)) \nonumber\\
     &\overset{(a)}{=}& D_a-m + (1-\lambda)f(a+1,D_a) + \lambda m -f(a+1,0)\nonumber\\
    &\overset{(b)}{=}& D_a-m + (1-\lambda)\left(\frac{D_a+a-\hat{J}^*}{\lambda} +m + \frac{1}{\lambda^2}\right) \nonumber\\
    && + \lambda m - a\hat{J}^* + \frac{a(a+1)}{2} \le 0,
\end{iarray}
where the equality $(a)$ follows from Proposition \ref{prop1} and \ref{prop2} and equality (b) is from \eqref{ds2}. We obtain the second condition:
\begin{iarray}
\label{cond2}
    D_a \ge   (1-\lambda+a\lambda)\hat{J}^* - a + 1 - \lambda \frac{a(a-1)}{2} -\frac{1}{\lambda}. 
\end{iarray}
Furthermore, for some state $(a,d)$ with $d>D_a$, based on Assumption \ref{ass2} and \eqref{c2go} the upper part which is associated with idling the terminal keeps increasing while the lower part stays the same, it is clear that 
\begin{iarray}
    \mu_0(a,d) - \mu_1(a,d) \ge 0,\,d>D_a,
\end{iarray}
once \eqref{cond1} and \eqref{cond2} are both satisfied.

Then the states are examined in which, based on $\pi_{\mathsf{D}}$, the terminal should be idle. Specifically, for some state $(a,D_a-1)$ with $a \ge D_1$, we obtain
\begin{iarray}
    && \mu_0(a,D_a-1) - \mu_1(a,D_a-1) \nonumber\\
    &=& D_a-1-m + (1-\lambda)(f(a+1,D_a-1)-f(a+1,0)) \nonumber\\
    && + \lambda(f(1,D_a-1+a)-f(1,a)) \nonumber\\
    &\overset{(a)}{=}& D_a-1-m + \frac{1-\lambda}{\lambda}(D_a-1) \le 0,
\end{iarray}
where the equality $(a)$ follows from \eqref{ds2}, and thus another condition is 
\begin{iarray}
\label{cond3}
    D_a \le \lambda m + 1.
\end{iarray}
With $1 \le a < D_1$,
\begin{iarray}
    && \mu_0(a,D_a-1) - \mu_1(a,D_a-1) \nonumber\\
    &=& D_a-1-m + (1-\lambda)(f(a+1,D_a-1)-f(a+1,0)) \nonumber\\
    && + \lambda(f(1,D_a-1+a)-f(1,a)) \nonumber\\
    &\overset{(a)}{=}& D_a-1-m + (1-\lambda)f(a+1,D_a-1) \nonumber\\
    && + \lambda m -f(a+1,0)\nonumber\\
    &\overset{(b)}{=}& D_a-1-m + (1-\lambda)\left(\frac{D_a+a-\hat{J}^*-1}{\lambda} +m + \frac{1}{\lambda^2}\right) \nonumber\\
    && + \lambda m - a\hat{J}^* + \frac{a(a+1)}{2} \le 0,
\end{iarray}
where the equality $(a)$ follows from Proposition \ref{prop1} and \ref{prop2} and equality (b) is from \eqref{ds2}. Hence, the following condition should be satisfied.
\begin{iarray}
\label{cond4}
    D_a \le   (1-\lambda+a\lambda)\hat{J}^* - a +2 - \lambda \frac{a(a-1)}{2} -\frac{1}{\lambda}. 
\end{iarray}
Following the same arguments before, for some state $(a,d)$ with $d<D_a-1$, based on Assumption \ref{ass2} and \eqref{c2go} the upper part which is associated with idling the terminal keeps decreasing when $d$ decreases while the lower part stays the same, it is thus clear that 
\begin{iarray}
    \mu_0(a,d) - \mu_1(a,d) \le 0,\,d<D_a-1,
\end{iarray}
once \eqref{cond3} and \eqref{cond4} are both satisfied. To summarize, the valid solution should satisfy the following conditions as well as \eqref{mjd}.
\begin{IEEEeqnarray}{rCll}
    && (1-\lambda+a\lambda)\hat{J}^* - a + 1 - \lambda \frac{a(a-1)}{2} -\frac{1}{\lambda} &\le D_a \le  ... \nonumber\\
    && (1-\lambda+a\lambda)\hat{J}^* - a +2 - \lambda \frac{a(a-1)}{2} -\frac{1}{\lambda},\,&\textrm{if }1 \le a < D_1; \nonumber\\
    && \lambda m \le D_a \le \lambda m +1,&\textrm{if } a \ge D_1.
\end{IEEEeqnarray}
Solving for the solution to the above, we obtain $D_1$ is the unique positive solution to the following equation:
\begin{iarray}
\label{2f}
    \frac{1}{2} D_1^2 + \left(\frac{1}{\lambda}-\frac{1}{2}\right)D_1-m = 0. \nonumber\\
\end{iarray}
The uniqueness of the positive solution can be shown based on the above formula. The rest of the thresholds are given by
\begin{iarray}
\label{da}
    D_a = \left\{\,
        \begin{IEEEeqnarraybox}[][c]{l?s}
        \IEEEstrut	
        \left(1-\lambda+a\lambda\right) D_1 -\lambda \frac{(a-1)a}{2}, &  if $1 \le a < D_1$;\\
    	\lambda m, & if $a \ge D_1$.
    	\IEEEstrut
    	\end{IEEEeqnarraybox}
    	\right.\nonumber\\
\end{iarray}
The average AoI is 
\begin{iarray}
\label{jj}
\hat{J}^* = \frac{1}{\lambda}+D_1.
\end{iarray}

Thus far, we have obtained the solution based on Assumption \ref{ass1} and \ref{ass2}. What is still left to do is to show that the solution is in consistency with both assumptions. Therefore, we first check the first assumption. The threshold policy nature has been validated in the preceding proofs, i.e., the optimal action given these differential cost-to-go functions is indeed threshold-based. Moreover, the monotonicity of the thresholds is shown as follows. Observe \eqref{da}, for $1\le a < D_1$
\begin{iarray}
\label{d_mono}
D_{a+1} - D_a = \lambda (D_1 -  a) > 0.
\end{iarray}
It follows that
\begin{iarray}
D_1 < D_2 <...<D_{D_1-1}.
\end{iarray}
Moreover, 
\begin{iarray}
\label{del}
D_{D_1-1} - D_{a,a \ge D_1} &\overset{(a)}{\le}&   D_1^2 -\lambda \frac{(D_1-1)D_1}{2} - \lambda m \nonumber\\
&\overset{(b)}{=}& \left(2-\frac{1}{\lambda}-\lambda\right)D_1 + \lambda -2 + \frac{1}{\lambda} \nonumber\\
&=&  0,
\end{iarray}
wherein the equality $(a)$ follows from the fact that $D_{D_1-1} \le D_{D_1}$ based on \eqref{d_mono}, and the equality $(b)$ is based on \eqref{2f} and massaging the terms. With this, we have shown that the obtained solution is consistent with Assumption \ref{ass1}.

For Assumption \ref{ass2} and specifically \eqref{mon}, when $a+d < D_1$,
\begin{iarray}
f(a,d+1) - f(a,d) \overset{(a)}{=} \hat{J}^* - a \overset{(b)}{>}  D_1 + \frac{1}{\lambda} -a >0, \nonumber\\
\end{iarray}
where the inequality $(a)$ follows from \eqref{ds3} and $(b)$ is from \eqref{jj}. When $D_1-a \le d < D_a$, the monotonicity is ensured by
\begin{iarray}
f(a,d+1) - f(a,d) \overset{(a)}{=} \frac{1}{\lambda} > 0, 
\end{iarray}
where the equality $(a)$ is based on Proposition \ref{prop1} and \eqref{ds2}. Note that the monotonicity around the connection point $(a,D_1-a)$ is intact by observing \eqref{mjd}. Furthermore, when $d \ge D_a$, $f(a,d)$ is a constant with respect to $d$ which is shown by Proposition \ref{prop2} and thus the monotonicity also holds. The connection point $(a,D_a)$ is validated by the same arguments as in \eqref{del}. We conclude the proof with this.

\section{Proof for Theorem \ref{thm2}}
\label{app_thm2}
Observe the thresholds in Theorem \eqref{thm1}. With $m=0$, the maximum threshold equals zero. Moreover, based on the monotonicity of the thresholds which is shown in the proof of Theorem \ref{thm1}, all thresholds are zero and hence the idle state space is an empty space. On the other hand, when $m$ goes to infinity, all the thresholds go to infinity with it and hence the idle state space approaches the entire space. Moreover, for any $m_1<m_2$, the thresholds associated with $m_1$ are no larger than those with $m_2$, and hence the monotonicity condition follows straightforwardly. This concludes the proof.
\bibliography{sm}
\bibliographystyle{IEEEtran}
\end{document}